# MID-INFRARED SPECTRA OF THE SHOCKED MURCHISON CM CHONDRITE: COMPARISON WITH ASTRONOMICAL OBSERVATIONS OF DUST IN DEBRIS DISKS

A. Morlok[a,b,f]

C. Koike[c]

N. Tomioka[a,d]

I. Mann[a,e]

K. Tomeoka[a]

[a]Department of Earth and Planetary Sciences, Faculty of Science, Kobe University, Nada, Kobe 657-8501, Japan, Phone: (+81) 788035743, Fax: (+81) 788035757, E-mail: tomeoka@kobe-u.ac.jp

[b]CRPG/CNRS, 54501 Vandoeuvre les Nancy, France, Phone: (+33) 3 83 59 42 11; Fax: (+33) 3 83 51 17 98,

[c]Department of Earth and Space Science, Graduate School of Science, Osaka University, Toyonaka, Osaka 560-0043, Japan. Phone: (+81) 66850 5803, Fax: (+81) 66850 5480, E-mail: koike-c@mua.biglobe.ne.jp

[d]Present address: Institute for Study of the Earth's Interior, Okayama University Misasa, Tottori 682-0193, Japan, E-mail: nao@misasa.okayama-u.ac.jp

[e]Present address: School of Science and Engineering, Kinki University, Kowakae 3-4-1, Higashi-Osaka, Osaka, 577-8502, Japan, Phone (+81) 667305403, Fax: (+81) 667272024 E-mail: mann@kindai.ac.jp

[f]Present address: Planetary and Space Sciences Research Institute, The Open University, Walton Hall, Milton Keynes MK7 6AA United Kingdom, Phone (+44) (0)1908 653219, Fax: +44 (0)1908 652559, E-Mail: a.morlok@open.ac.uk

**39 Pages,** 6 Figures, 2 Tables

Running Head:

Mid-infrared Spectroscopy of Shocked Murchison CM2

Editorial correspondence and proofs should be directed to

Andreas Morlok

Planetary and Space Sciences Research Institute

The Open University

Walton Hall

Milton Keynes

MK7 7NG

United Kingdom

Phone (+44) (0)1908 653219

Fax: +44 (0)1908 652559

E-Mail: a.morlok@open.ac.uk

**Abstract**

We present laboratory mid-infrared transmission/absorption spectra obtained from matrix of the hydrated Murchison CM meteorite experimentally shocked at peak pressures of 10–49 GPa, and compare them to astronomical observations of circumstellar dust in different stages of the formation of planetary systems. The laboratory spectra of the Murchison samples exhibit characteristic changes in the infrared features. A weakly shocked sample (shocked at 10 GPa) shows almost no changes from the unshocked sample dominated by hydrous silicate (serpentine). Moderately shocked samples (21–34 GPa) have typical serpentine features gradually replaced by bands of amorphous material and olivine with increasing shock pressure. A strongly shocked sample (36 GPa) shows major changes due to decomposition of the serpentine and due to devolatilization. A shock melted sample (49 GPa) shows features of olivine recrystallized from melted material.

The infrared spectra of the shocked Murchison samples show similarities to astronomical spectra of dust in various young stellar objects and debris disks. The spectra of highly shocked Murchison samples (36 and 49 GPa) are similar to those of dust in the debris disks of HD113766 and HD69830, and the transitional disk of HD100546. The moderately shocked samples (21–34 GPa) exhibit spectra similar to those of dust in the debris disks of Beta Pictoris and BD+20307, and the transitional disk of GM Aur. An average of the spectra of all Murchison samples (0–49 GPa) has a similarity to the spectrum of the older protoplanetary disk of SU Auriga. In the gas-rich transitional and protoplanetary disks, the abundances of amorphous silicates and gases have widely been considered to be a primary property. However, our study suggests that impact processing may play a significant role in generating secondary amorphous silicates and gases in those disks. Infrared spectra of the shocked Murchison samples also show similarities to the dust from comets (C/2002 V1, C/2001 RX14, 9P/Tempel 1, and Hale Bopp), suggesting that the comets also contain shocked Murchison-like material.

1. **Introduction**

Silicates are among the most common constituents of dust in space and they change their composition and structure during the different stages of evolution from interstellar dust to dust in planetary systems (e.g., Molster and Waters, 2003). Those evolutional changes of silicates can be traced by studying characteristic emission features detected in astronomical observations and comparing them to laboratory measurements (e.g., Colangeli et al., 2003). Recently improved space- and ground-based telescope facilities provide a growing number of astronomical infrared spectral observations of young stellar objects and debris disks that are evolving toward planetary systems (e.g., references in Table 1), and those observations allow us to deduce the mineralogical nature of dust and physical processes taking place in the forming planetary systems.

Most of previous astronomical infrared studies of the evolution of silicates have focused on the physical processes such as vaporization, condensation and annealing in the interstellar medium and in the early solar nebula. However, when we consider the evolutional changes of silicates in planetary systems, there is another important physical process that should be considered, that is, shock. During the formation of dust by planetesimal collisions, the dust would undergo significant shock processes and thus experience significant changes in mineralogical properties. It is, therefore, important to understand how the shock affects the mineralogical properties of the planetesimal components, and how those changes affect the infrared spectroscopic properties.

In order to interpret astronomical infrared data, laboratory spectroscopic measurements of dust analogs are necessary. Previously, such studies mainly focused on terrestrial and synthetic minerals (e.g., Henning and Mutschke, 1997; Hofmeister, 1997; Bowey and Hofmeister, 2005; Chihara et al., 2007). However, since planetesimals around the forming and evolving stars probably consist of materials similar to those in our early solar system, infrared measurements of meteorites would provide a more important source of information.

CM type carbonaceous chondrites can be regarded to be a typical analogue of hydrated asteroids. Reflectance spectroscopic measurements indicate that 30–40 % of known asteroids are C-type, which many workers believe to be the source of the hydrated carbonaceous chondrites, such as CM-CI chondrites, as well as hydrous interplanetary dust particles (e.g., Rivkin et al., 2002; Jones et al., 1990). Previous shock experiments suggest that hydrated asteroids produce dust particles during collisions at a much higher rate than anhydrous asteroids, because of their high $H_2O$ contents and porosities (Tomeoka et al., 2003). Thus, CM chondrites likely represent a typical source of dust in our solar system and a good analogue for dust in other circumstellar systems.

Here we present the results of mid-infrared spectroscopic measurements of matrix of the experimentally shocked Murchison CM chondrite. Our study focuses on the matrix of Murchison, because it is composed of a porous aggregate of fine-grained hydrous minerals and thus probably represents the main source of dust (Tomeoka et al., 2003; Flynn and Durda, 2004). The Murchison samples used in our study have been recovered from the previous shock experiments (Tomeoka et al., 1999), in which the samples were shocked at peak pressures from 4 GPa to 49 GPa, and the recovered samples were studied using various microscope techniques. More recently, the matrix mineralogy of the recovered samples were studied in detail using transmission electron microscopy (Tomioka et al., 2007). Recent theoretical modeling of impact shock pressures for main belt asteroidal collisions showed that generated pressures are in the range below 50 GPa (Nakamura and Michel, 2008), which overlaps very well with the pressure range dealt with in our study. Our goal is (1) to examine how impacts affect the infrared properties of the Murchison matrix, (2) to compare infrared spectra of the shocked Murchison matrix with infrared spectra of dust obtained from selected observations of young stellar objects, debris disks, and solar system comets, and (3) to deduce the mineralogical nature of circumstellar dust in various stages of the formation of planetary systems.

## 2. Samples and Techniques

The previous shock experiments on Murchison (Tomeoka et al., 1999) were performed using a single stage propellant gun at the National Institute for Materials Science in Japan. The Murchison samples were shocked in nine experiments at peak pressures of 4, 10, 21, 26, 28, 30, 34, 36, and 49 GPa. The targets were disks of Murchison, 5.7−10.0 mm in diameter and 1.0−3.5 mm in thickness, encapsulated in stainless steel containers. The projectiles, which were made of stainless steel, aluminum or molybdenum, were shot at the targets with velocities ranging from 0.50 to 1.80 km/s. Further details about the shock experiments are described in Tomeoka et al. (1999). The Murchison samples shocked in eight experiments at 10, 21, 26, 28, 30, 34, 36, and 49 GPa were subjected to the present IR measurements.

After areas of meteorite matrix on polished thin sections were identified using backscattered electron imaging, matrix materials were separated using a tungsten needle. The separated samples were crushed to very fine grain size (<1 μm) using a diamond compression cell, which allows us to analyze the samples *in situ*, through a diamond window using an infrared microscope. This technique is known to produce results similar to those obtained using KBr pellets, but there might be a small medium effect resulting in small peak shifts (see Morlok et al. (2006) and Osawa et al. (2001) for details).

Mid-infrared transmission/absorption spectra were measured from 2.5 μm to 23.0 μm with a 0.1 μm (4 $cm^{-1}$) spectral resolution using a Nexus 670 FTIR infrared microscope at Kyoto Pharmaceutical/Osaka University. The spectrum presented here for a sample shocked at each shock pressure is an average of spectra measured from several aliquots, each taken from different locations of the meteorite matrix. Each single measurement consists of 256 added scans. The results are presented in absorbance, A (A = ln(100/T), with T being the measured transmission, normalized to the strongest silicate feature at ~10 μm. However, the mineral phase responsible for the strongest feature changes with increasing shock pressure—from serpentine to glass to olivine. These minerals have different absorbance coefficients, and thus, the normalization may affect the relative intensities of the features; this possibility should be kept in mind when intensities of features between spectra of different samples are compared.

Since the shocked samples are very fragile, they were embedded in resin for handling. To avoid contributions from the resin, we measured spectra for the resin and subtracted its bands from each spectrum.

### 3. Mineralogy of Natural and Experimentally Shocked Murchison Matrix

To help readers understand the following descriptions, we here provide a brief summary of the mineralogy of natural Murchison and the results of the previous TEM study on the experimentally shocked Murchison samples (Tomioka et al., 2007).

Natural Murchison contains chondrules of 100–500 μm in diameter embedded in a high volume fraction (~64 %, data from McSween (1979)) of matrix. The porosity of bulk Murchison has been reported to be 23 % (Corrigan et al., 1997). The matrix consists mainly of fine-grained hydrous minerals, mostly Fe-Mg serpentine [$(Fe^{2+}Mg)_6(Fe^{3+}Si)_4O_{10}(OH)_8$] and tochilinite [$6FeS \cdot 5Fe(OH)_2$] (Barber 1981; Tomeoka and Buseck 1985), that account for most of the water (~12 wt%, data from Fuchs et al. (1973)) in this meteorite. Minor constituents in matrix are magnetite, troilite, pentlandite, chromite, calcite, and organic compounds.

In the Murchison matrix sample experimentally shocked at 10 GPa, tochilinite is completely decomposed to an amorphous material. Fe-Mg serpentine is partially decomposed to Si-rich glass and decreases in amount with increasing pressure from 10 GPa to 30 GPa and is completely decomposed at 36 GPa. At 49 GPa, the matrix is extensively melted and consists mostly of fine-grained aggregates of Fe-rich olivine and less abundant low-Ca pyroxene embedded in Si-rich glass. The aggregates of olivine and pyroxene appear to have crystallized from the matrix melt. The major constituent minerals of the Murchison matrix shocked at the different pressures are summarized in Table 2.

## 4. Results

The spectrum of the unshocked Murchison matrix sample in the band range below 15 μm shows a general similarity to that of the dominating phase, serpentine (Fig. 1).

A sharp feature at 9.9−10.2 μm is the strongest band in the spectra of the unshocked sample and all samples shocked at pressures up to 34 GPa (Fig. 1). The band position shifts gradually from 10.2 μm to 9.9 μm with increasing pressure up to 34 GPa, and along with this shift, the band shape becomes sharper. These changes were probably caused by the breakdown of serpentine and tochilinite in the original matrix to Si-rich glass**.**

A second characteristic feature appears as a shoulder at 11.2 μm at 10 GPa. This feature gradually increases in intensity with increasing pressure and turns into the strongest band at 11.3 μm at 36 GPa and 49 GPa. The increasing relative strength of this band probably reflects the increasing amounts of olivine in the matrix. The same trend is confirmed **by the presence of** other characteristic olivine bands. For example, a broad olivine feature appears between 19.5 μm and 20.0 μm at 21 GPa, and its intensity increases with increasing pressure up to 49 GPa. A strong, broad band at 10.5 μm that appears at 36 GPa could be of an amorphous material with olivine composition — either highly shocked olivine or quenched melt (Koike and Tsuchiyama, 1992) .

Water bands typical for serpentine (Osawa et al., 2001; Hofmeister and Bowey, 2006) occur between 6.0–6.3 μm at pressures from 0 GPa (unshocked) to 34 GPa (Fig. 1), and disappear at higher pressures. Of other typical water features for serpentine in the 2.9–3.0 μm region, dominating is a broad feature at ~3 μm (Figs. 2a, b). This band is characteristic of $H_2O$ or stretching modes of OH not fixed in the crystal structure (Osawa et al., 2001; Hofmeister and Bowey, 2006). The intensity of this band decreases steadily with increasing pressure up to 30 GPa except for 26 GPa, at which the intensity of this band is as high as for the unshocked sample. This erratic behavior of the 26 GPa sample may be ascribed to uneven distribution of the shock effects in the sample. The shape and the position of the ~3 μm band are nearly constant up to 30 GPa (Figs. 2a, b). However, this feature exhibits a significant drop in intensity at each step from 30 GPa to 34 GPa, from 34 GPa to 36 GPa, and finally from 36 GPa to 49 GPa. In addition to these changes in intensity, the band position slightly shifts from 2.95 μm (unshocked to 30 GPa) to 3.03 μm (36 GPa) and back to 2.94 μm (49 GPa) (Figs. 2a, b).

An additional OH feature expected to be present at ~2.7 μm (Osawa et al., 2001; Hofmeister and Bowey, 2006) is visible only as a shoulder (due to the low resolution of 4 cm$^{-1}$) at 2.9 μm at 0–30 GPa (Figs. 2a, b). However, this feature gradually decreases in intensity with increasing pressure, and disappears at 34 GPa (with the 26 GPa sample as an exception). The intensity of a broad serpentine OH band between 15.4 μm and 16.4 μm decreases with increasing pressure from unshocked to 30 GPa (Fig. 1).

A carbonate band at 7.0 μm appears in all spectra except 36 GPa (Fig. 1). A minor carbonate band at ~11 μm is probably too weak to significantly interfere with the emerging olivine band at 11.3 μm. Another minor carbonate band at ~14 μm is missing, which could be due to the relatively low resolution of the laboratory spectra (Koike et al., 2003a). The absence of the 7.0 μm band in the 36 GPa spectrum is probably a result of sampling bias, since carbonates are relatively minor in abundance in the Murchison matrix.

Minor phases such as pyroxene, magnetite and sulfides are not visible in the spectra, which is probably mainly due to their low abundances. In the case of sulfides, it may also be due to lack of significant spectral features in the observed wavelength range; for example, FeS shows strong IR bands only at >23 μm (Hony et al., 2002).

## 5. Discussion

The results of our study show that the systematic changes in the infrared spectral features of the Murchison samples shocked at 0–49 GPa are consistent with the changes in mineralogy observed in the earlier TEM study (Tomioka et al., 2007). We classified the spectra into four groups I – IV based on their major spectral features at each shock stage (Table 2).

### 5.1. Water Bands

The water bands characteristic of serpentine reveal especially distinct changes. The intensity of the water band at ~3 μm shows only a small, steady decrease at pressures from 0 GPa (unshocked) to 30 GPa (Figs. 2a, b). This confirms the results of TEM studies that show a gradual decrease in hydrous serpentine and a simultaneous increase in anhydrous phases in this pressure range. The sharp drop in the intensity of the ~3 μm band from 30 GPa to 34/36 GPa (Fig. 2a) as well as the disappearance of the 6.0–6.3 μm band at 36 GPa and higher pressures (Fig. 1) indicates that major volatilization took place in this pressure range. This is consistent with the mineralogical observation showing that the complete decomposition of serpentine took place at 36 GPa. A further drop in the intensity between 36 GPa and 49 GPa is also consistent with extensive melting of the meteorite matrix which occurred at 49 GPa.

Minor bands characteristic of water were observed even in the spectrum of the Murchison sample shocked at 49 GPa. Because we cannot expect any surviving hydrous minerals in this sample, these bands probably resulted from water condensed from vapor, which had been volatilized from hydrous phases during impact, and adsorbed on the fine-grained minerals in the sample container.

*5.2. Comparison to astronomical observations*

We here compare the infrared spectra of the shocked Murchison samples with astronomical infrared spectra of selected young stellar objects, debris disks, and solar system comets. Dust properties reported for the astronomical objects that we used in this study are listed in Table 1. Our comparison is focused on the spectral range from 5 μm to 23 μm. Some of the astronomical spectra are limited to the narrow atmospheric window from 8 μm to 13 μm.

In the following comparison, we mainly focus on the major spectral features indicating the composition and mineralogy of the observed dust and samples. However, one has to keep in mind that infrared features do not provide full information about the composition and mineralogy of the observed materials and there is always a certain degree of ambiguity in their interpretation. Laboratory measurements and model calculations of optical properties indicate that infrared features also depend on various other factors including size, shape, structure, and temperature of the observed materials (e.g., Bowey et al., 2001; Min et al., 2005; Mann et al., 2006). In the following discussion, we will not go into details of those other factors.

## 5.2.1. Debris Disks

Debris disks constitute a phase in the formation of a planetary system following the protoplanetary disk phase. In the debris disk phase, the vast majority of the gases is removed (e.g., Hollenbach et al., 2000), and the dust is replenished by collisional destruction of planetesimals (e.g., Wyatt, 2008).

The sun-like main-sequence star BD+20307 has an infrared excess due to warm dust in a distance of ~1 AU from the central star (Song et al., 2005). Dust in this 300 Myr old system has been assumed to be similar to the dust generating the zodiacal light in our solar system (Song et al., 2005). Its mid-infrared spectrum shows a strong, broad band at 9.8 μm and a weaker, broad shoulder-like band at 11.0 μm (Fig. 3). These features were interpreted as a result of the mixing of a crystalline material and amorphous, interstellar-medium like silicates (Song et al., 2005). The spectrum of the Murchison sample shocked at 34 GPa has maxima at 9.9 μm and 11.2 μm. A minor shift of 0.1 μm in band positions results in features nearly identical to those of BD+20307 (Fig. 3).

A similar mid-infrared spectrum was observed for the 10–20 Myr old nearby main-sequence (A5V) star Beta Pictoris (Telesco et al., 2005). The edge-on debris disk of this star shows asymmetries, which were interpreted as either dust rings formed by recent collisional disruptions of planetesimals (Telesco et al., 2005) or dust formed from such events kept in resonance trappings caused by already formed planets (Wyatt et al., 2006). High-resolution mid-infrared observations of the central part of this system (<3.2 AU) show two major features at 10.1 μm and 11.1 μm, although the former is degraded by the effects of atmospheric ozone (Fig. 3; Okamoto et al., 2004). Together with smaller bands, these features were interpreted as a result of the mixing of both crystalline and amorphous olivine, with grain size of <2 μm. The spectrum of the Murchison sample shocked at 34 GPa provides a good fit in both band positions and shapes, with strong features at 9.9 μm and 11.2 μm. However, we note that a SPITZER IRS spectrum of Beta Pictoris reported by Chen et al. (2007) shows bands considerably different from those reported by Okamoto et al. (2004) (Fig. 3).

HD113766 is a 16 Myr old binary F3/F5V system with a debris disk (Chen et al., 2006). The SPITZER spectrum of this star shows strong, sharp bands at 10.0 μm, 11.1 μm, and broader bands at 16.3 μm and 19.0 μm (Fig. 3). The spectral fitting analysis by Chen et al. (2006) suggests that the dust is a mixture of mainly amorphous olivine and pyroxene-like material, and a minor amount of amorphous carbon. Lisse et al. (2008) suggested that forsterite and sulfides are dominating minerals. The spectrum of the Murchison sample shocked at 49 GPa shows a similarity to that of HD113766 with strong bands at 10.2 μm and 11.3 μm (Fig. 3). However, for a better fit, a slight shift of 0.2–0.8 μm in the band positions is necessary. Since band positions of olivine features are known to shift to longer wave lengths with increasing Fe/Mg ratios (Koike et al., 2003b), the slight differences in the band positions between these two spectra may reflect difference in Fe/Mg ratio in olivine.

HD69830 is an old, 3–10 Gy old system (Lisse et al., 2007a). The dust emission around this K0V star, which is ~1000 times as intense as the zodiacal light in our solar system, is probably generated from part of a dust ring at a distance of ~1 AU from the central star (Lisse et al., 2007a). The spectrum of this star has an overall similarity to that of HD113766, with strong features at 10.9 μm, 16.6 μm, and 19.1 μm, and weaker features at 10.3 μm and 11.9 μm (Fig. 3). The spectral fitting analysis indicates that the dust is a mixture of amorphous carbon, olivine and pyroxene, with some carbonates (Lisse et al., 2007a). The spectrum of the Murchison sample shocked at 49 GPa is similar to that of HD69830 (Fig. 3). A better fit can be obtained by a calculated mixture of 2/3 of the 36 GPa spectrum and 1/3 of the 49 GPa spectrum. Remaining slight differences could be explained by difference in Fe/Mg ratio of olivine, similarly to the case of HD113766.

In summary, the mid-infrared spectra of the Murchison samples shocked to moderate to very high shock stages (34–49 GPa) show similarities to the astronomical spectra obtained from second-generation dust produced by collisions of planetesimals in the four debris disks. Especially the 9–12 μm silicate features of the shocked material provide good fits.

### 5.2.2. Protoplanetary and Transitional Disks

Transitional disks constitute a phase in the transition from a gas rich protoplanetary disk to a debris disk, thus representing a more primordial stage than the debris disks in the disk evolution. The optically thin parts of the disks have already been cleared of gas, while optically thick parts are still gas rich. The clearing of gas may have resulted from accretion by planets (Calvet et al., 2005).

GM Aur is a young Tauri star in the transitional phase, in which part of the disk at a distance of 5–24 AU from the central star may have been cleared by planet formation (Calvet et al., 2005). The spectrum from this star has a broad feature at 9.9 μm and a weaker feature at 11.3 μm (Fig. 4). An average of the spectra of the Murchison samples shocked at 21–34 GPa, with the same weight, provides a good fit in both band shapes and positions.

HD100546 is a ~10 Myr old pre-main sequence Herbig AE star, in which a collisional cascade triggered by giant planet formation was modeled as a source for the dust in this system (Bouwman et al., 2003). The mid-infrared spectrum from this star shows a strong, sharp band at 11.3 μm and weaker features at 10.1 μm, 11.1 μm, and 11.9 μm (Fig. 4). Further strong bands are found at 16.3 μm and 19.7 μm (Sloan et al., 2003; van Boekel et al., 2005). Calculations of mineralogical compositions of the dust suggest that it may consist of amorphous silicates and carbonaceous material with minor amounts of forsteritic olivine, water ice, iron (Bouwman et al., 2003), or amorphous material with olivine composition and minor crystalline forsterite, enstatite and silica (van Boekel et al., 2005), or forsterite, amorphous and crystalline pyroxene, Mg-carbonate, sulfide, $H_2O$, and carbon (Lisse et al., 2007b). We found that a calculated mixture of 2/3 of the 36 GPa spectrum and 1/3 of the 49 GPa spectrum yields a similar spectrum with a band at 11.3 μm and flat bands at 10.2–10.4 μm and 19.5 μm (Fig. 4).

Protoplanetary disk SU Auriga is a ~9 Myr old star in its Tauri stage (Furlan et al., 2006). The mid-infrared spectrum is dominated by a broad feature at 9.9 μm and a small sharp feature at 11.3 μm (Fig. 5). The spectral fitting analysis suggests that the dust may consist mainly of amorphous and minor crystalline olivine-like material (Honda et al., 2006). A calculated average of the spectra from all Murchison samples (from unshocked to 49 GPa), with the same weight, provides a similar spectrum with bands at 10.1 μm and 11.3 μm. Although the carbonate band at ~7 μm is missing in the astronomical spectrum, a potential water feature is found at 6.3 μm (Fig. 5). These results suggest that this protoplanetary disk consists of a mixture of the materials having experienced a wide range of shock metamorphism, from unshocked to shock melted stage.

In circumstellar disks of the Herbig AE and Tauri types, crystallization of the amorphous, starting material by annealing has been considered as an indication of increasing processing of material (e.g., van Boekel et al, 2005; Kessler-Silacci et al., 2006; Murata et al., 2007). However, the previous TEM (Tomioka et al., 2007) and the present infrared studies of the shocked Murchison samples reveal that major amounts of amorphous silicates are produced during secondary impact processes. Thus, planetesimal collisions in the circumstellar disks could also be responsible for the formation of amorphous material and affect the ratio of amorphous/crystalline materials in those disks.

Our study also reveals that dramatic volatilization occurred during impacts of the volatile-rich Murchison samples. In the Murchison shock experiments, the predominant volatilized component is $H_2O$. However, in the debris disks, impacts could occur at much higher velocities than those in the shock experiments, and in that case, heavier components would also be vaporized and gases of many species could be generated from the dust and planetesimals (Czechowski and Mann, 2007). In the absence of an efficient process to remove those gas components, we envision that the amounts of gases would possibly become sufficiently large and influence the evolution of the debris disks.

*5.2.3. Comets*

Oort cloud comets C/2002 V1 and C/2001 RX14 were observed using the COMIC/Subaru spectrometer (Honda et al., 2004). The observations indicate that the comets consist mainly of amorphous olivine and pyroxene, with a minor amount of crystalline olivine. The broad feature at 10.2 μm and the small feature at 11.3 μm in the C/2002 V1 spectrum and the broad features at ~10 μm and 11.4 μm in the C/2001 RX14 spectrum can be reproduced by averaging the spectra from all Murchison samples (unshocked to 49 GPa) (Fig. 6), similarly to the case of SU Auriga described above.

The mineralogical composition of comet 9P/Tempel 1 was deduced based on the data obtained by space based spectroscopy (characteristic bands at 10.2 and 11.2 μm) (Lisse et al., 2007b) and ground based spectroscopy (characteristic bands at 9.9 and 10.8– 11.3 μm) (Harker et al., 2007) of the ejecta produced by the impact of a probe on the comet (Fig. 6). Lisse et al. (2007b) reported that the **cometary** dust is dominated by crystalline silicates (mainly forsteritic olivine and low-Ca pyroxene) and sulfides, with minor amounts of carbonates, phyllosilicate, and amorphous silicates, whereas Harker et al. (2007) reported that the cometary ejecta consist mainly of amorphous silicates and carbon, with minor amounts of crystalline olivine.

The band positions of the two major bands in the spectra obtained by these two groups are similar, but their relative intensities are different. An average of the spectra from all Murchison samples provides a better fit with the comet spectrum obtained by Harker et al. (2007), with peaks at 10.1 μm and 11.3 μm, than with the comet spectrum obtained by Lisse et al. (2007b) (Fig. 6). However, as mentioned earlier, there is always a certain ambiguity in interpretation of infrared features. Thus we note that more information is necessary to provide a clearer identification about the mineralogy of the cometary dust.

Comet Hale Bopp shows a mineralogical composition dominated by crystalline (forsteritic) olivine, pyroxene, carbonate, sulfide, and amorphous carbon (Lisse et al., 2007b). The spectrum of this comet is characterized by strong bands at 11.3 μm, 19.3 μm, and ~24.0 μm and weaker bands at 11.9 μm and 16.4 μm (Fig. 6). A calculated mixture of 2/3 of the 36 spectrum and 1/3 of the 49 GPa spectrum provides a good fit with strong features at 11.3 μm and 19.5 μm.

While comets are generally regarded as being composed of pristine interstellar material, there is growing evidence that their materials have been affected by various alteration events, including hydration (Lodders and Osborne, 1999) and collisional processing (Lederer et al., 2008; Lisse et al., 2007b). The recent laboratory analyses of cometary particles returned by the Stardust mission provide abundant data indicating that they are similar to the carbonaceous chondrites (e.g., Zolensky et al., 2006). Although there has been no report of hydration in the cometary particles returned by Stardust, a TEM study reports evidence suggesting that some of the cometary particles experienced hypervelocity impacts before capture (Tomeoka et al., 2008). The results of our study suggest that planetesimal collisions may have already occurred during the protoplanetary disk stage, and materials that had gone through such events may have been preserved in comets.

## 6. Summary

The mid-infrared measurements of matrix of the experimentally shocked Murchison meteorite show characteristic changes in the infrared features that are consistent with the mineralogy observed in the earlier TEM study.

While the shock at a pressure of 10 GPa results almost no changes in the infrared spectrum, typical phyllosilicate features are gradually replaced by bands of amorphous material and olivine in the spectra of the samples shocked at moderate pressures from 21 GPa to 34 GPa. Drastic changes in the infrared features due to volatilization are observed at 36 GPa, where hydrous serpentine was completely decomposed to an amorphous material. Further changes due to volatilization are observed at 49 GPa, where the sample was subjected to extensive melting.

The infrared spectra of the shocked Murchison samples show similarities to the astronomical spectra of dust in various young stellar objects and debris disks. The spectrum of the moderately shocked Murchison sample (34 GPa) is similar to those of dust in the debris disks of Beta Pictoris and BD+20307. The spectra of the highly shocked Murchison samples (36 GPa and 49 GPa) are similar to those of another olivine-rich debris disks of HD 113766 and HD 69830.

The moderately shocked (21–34 GPa) and highly shocked (36 and 49 GPa) Murchison samples exhibit spectra similar to dust in the transitional disks of GM Aur and HD100546, respectively. An average of the spectra of all Murchison samples (0–49 GPa) has a similarity to the spectrum of the older protoplanetary disk of SU Auriga. In these young stellar objects and debris disks, the abundances of amorphous silicates and gases have widely been considered to be a primary property. However, our study suggests that impact processing may play a significant role in generating secondary amorphous silicates and gases in those disks.

Infrared spectra of the shocked Murchison samples also show similarities to those of dust from comets (C2002 V1, C/2001 RX14, 9P/Tempel 1, and Hale Bopp), suggesting that the comets also contain Murchison-like material that has undergone impact processing.

## Acknowledgements

This work was supported by "The 21st Century COE program of Origin and Evolution of Planetary Systems" of the Ministry of Education, Culture, Sports, Science and Technology, and the Grant-in-Aid for Scientific Research to A. Morlok, N. Tomioka, (No. 17740350), K. Tomeoka (No. 16204042), and I. Mann (No. 19540475). I.M. thanks the Open Research Center at Kinki University for support.

Many thanks for providing the astronomical spectra of this paper to C.Chen for Beta Pictoris, to Y.K.Okamoto (Beta Pictoris), I.Song (BD+20307), C.Lisse (HD113766, HD69830, Hale Bopp and 9P/Tempel 1), E.Furlan (GM Aur, SU Auriga), D.E. Harker (9P Tempel 1) and M.Honda (C/2002 V1, C/2001 RX14).

This work is partially based on observations made with the *Spitzer Space Telescope*, which is operated by the Jet Propulsion Laboratory, California Institute of Technology, under NASA contract 1407.

Part of this work is based on data collected at the *SUBARU* telescope, which is operated by the National Astronomical Observatory of Japan.

Part of this work is based on observations obtained at the Gemini and Keck Observatories. The Gemini Observatory is operated by the Association of Universities in Astronomy, Inc., under a cooperative agreement with the NSF on behalf of the Gemini Partnership.

Part of this work is based on observations with ISO, an ESA project with instruments funded by ESA Member States (especially the PI countries: France, Germany, the Netherlands, and the United Kingdom) and with the participation of ISAS and NASA.

Terrestrial serpentine was provided by Y. Wada (Kobe University). Also many thanks to A. Nakamura (Kobe University) and M. Köhler (Columbia University) for contribution to the manuscript. This research has made use of the SIMBAD database, operated at CDS, Strasbourg, France.

Table 1:  Dust properties in the astronomical objects used in this study (literature data) and pressures at which the Murchison samples showing best fit spectra were shocked (shown in GPa). $M_o$ = Mass of our sun. Sources**:** Sources**:** 1: Furlan et al., 2006 2: Calvet et al., 2005 3: Watson et al., 2007 4: Schegerer et al., 2006 5: Lisse et al., 2007a  6: van Boekel et al., 2005 7: Okamoto et al., 2004 8: Chen et al., 2007 9: Song et al., 2005 10: Chen et al., 2006 11: Lisse et al., 2008 12: Sackmann, 1993 13: Harker et al., 2007 14: Lisse et al., 2007b 15: Honda et al., 2004.

| | Spectral class | Dust temperature (K) | Age (Myr) | Stellar photospheric temperature (K) | Mass star (in $M_o$) | Disk stage | Distance Sun – Dust (AU) | Best fit Murchison sample (GPa) | Sources |
|---|---|---|---|---|---|---|---|---|---|
| SU AUR | G1 | | 8.7 | 5945 | | Protoplanetary disk | | 0–49 | 1, 2 |
| GM Aur | K5 | 403 | 1.3–1.8 | 4060 | | Transitional disk | | 21–34 | 3, 4 |
| HD100546 | Be9V | 250–410 | <10 | 4021 | 2.4 | Transitional disk | 12–14 | 36–49 | 5, 6 |
| Beta Pictoris | A5V | <120 (> 20AU) 400-600 (<10 AU) | 12–20 | 8000–8200 | 1.8 | Debris disk | Center (<=3.2) | 34 | 7, 8 |
| BD+20307 | G0 | 650 | 300 | 6000 | ~1 | Debris disk | 1 | 34 | 9 |
| HD113766 | F3/F5V | 440 | 16 | 6870 | 1.8 | Debris disk | 1.8 | 49 | 10, 11 |
| HD69830 | K0V | 340–410 | 3000–10000 | 5385 | 0.86 | Debris disk | 0.9–1.2 | 36–49 | 5 |
| Sun | G2V | | 4.56 | 5778 | 1 | Evolved debris disk | | | 12 |
| 9P/Tempe1 | | 190–410 | | | | Comet (Short period) | 1.5 | 0–49 | 13, 14 |
| C/2002 V1 | | 280–295 | | | | Comet (Long period) | 1.2 | 0–49 | 15, 16 |
| C/2001 RX 14 | | 245 | | | | Comet (Long period) | 2.1 | 0–49 | 11 |
| Hale Bopp | | 190–410 | | | | Comet (Long period) | 2.8 | 36–49 | 14 |

**Table 2: Summary of strong, characteristic spectral features of the Murchison matrix samples as observed by FTIR in this study, divided into four shock stages. For comparison, the mineralogy observed by TEM (Tomioka et al., 2007) is shown.**

| Shock stage | Pressure (GPa) | Mineralogy (TEM, Tomioka et al., 2007) | Mineralogy (FTIR, this study) | Characteristic spectral features (μm) | $H_2O$ features (μm) |
|---|---|---|---|---|---|
| I. Unshocked/Weakly shocked | 0–10 | **Si-rich glass Serpentine** | Phyllosilicates<br><br>Carbonates | 10.0–10.2<br>15.4–16.2<br>7.0 | 6.1–6.2<br>2.9–3.0 |
| II. Moderately shocked | 21–34 | **Si-rich glass Serpentine Olivine (Fo$_{98-99}$) Low-Ca pyroxene (En$_{92-98}$) Magnetite Fe sulfide** | Mixture of phyllosilicates and amorphous silicates<br>Olivine<br>Carbonates | 9.8–9.9<br>16.3–16.4<br><br>11.2<br>6.9–7.0 | 6.0–6.3<br>3.0 |
| III. Strongly shocked | 36 | **Si-rich glass Olivine (Fo$_{90-98}$) Low-Ca pyroxene (En$_{72-97}$) Magnetite Fe sulfide** | Mixture of olivine and amorphous silicates | 10.4, 10.7, 11.3, 19.5 | 3.0 |
| IV. Shock melted | 49 | **Olivine (Fo$_{40-69}$) Low-Ca pyroxene (En$_{72-77}$) Si-rich glass Magnetite** | Olivine<br><br>Carbonates | 10.2, 11.3, 12.0, 19.6<br><br>7.0 | 2.9 (weak) |

**Figure Captions:**

Fig. 1: 5 μm to 23 μm infrared spectra of an unshocked Murchison matrix sample (0 GPa) and shocked Murchison matrix samples (10−49 GPa). Shown on the top by a dotted line is a spectrum of terrestrial serpentine, representing the dominating phase in the Murchison matrix. Bands of characteristic minerals are shown by vertical broken lines. The sharp band at 15 μm is a contribution from atmospheric $CO_2$.

Fig. 2: (a) 2.5 μm to 4 μm infrared spectra of unshocked (0 GPa) and shocked (10−49 GPa) Murchison matrix samples, showing water bands. The bands in the 3.3−3.6 μm range are a contribution from $CH_2$ that is probably from the organic components in the matrix (b) 2.5 μm to 4 μm infrared spectra normalized to the same intensity, to allow a better comparison of changes in band shape.

Fig. 3: Comparison of astronomical infrared spectra of dust in selected debris disks to laboratory infrared spectra of shocked Murchison matrix samples. Top: Beta Pictoris and BD+20307 *vs.* Murchison shocked at 34 GPa. The spectrum from Okamoto et al. (2004) was taken from the central part of Beta Pictoris (<3.2 AU). Middle: HD113766 *vs.* Murchison shocked at 49 GPa. Bottom: HD69830 *vs.* Murchison shocked at 36 and 49 GPa (a mixture of 2/3 of the 36 GPa spectrum and 1/3 of the 49 GPa spectrum). Vertical broken lines indicate olivine features. The intensities of spectra are presented in linear arbitrary units.

Fig. 4: Comparison of spectra of dust in selected transitional disks to spectra of shocked Murchison matrix samples. Top: GM Aur *vs.* Murchison shocked at 21−34 GPa (an average of the 21−34 GPa spectra). Bottom: HD100546 *vs.* Murchison shocked at 36 and 49 GPa (a mixture of 2/3 of the 36 GPa spectrum and 1/3 of the 49 GPa spectrum). Vertical broken lines indicate olivine features.

Fig. 5: Comparison of a spectrum of dust in the protoplanetary disk of SU Auriga to a spectrum obtained by averaging the spectra of Murchison matrix samples shocked at 0−49 GPa. Vertical broken lines indicate olivine features.

Fig. 6: Comparison of spectra of selected comets to spectra of shocked Murchison matrix samples. Top: Hale Bopp *vs.* Murchison shocked at 36 and 49 GPa (a mixture of 2/3 of the 36 GPa spectrum and 1/3 of the 49 GPa spectrum). Bottom: 9P/Tempel 1, C/2002 V1 and C/2001 RX14 *vs.* Murchison shocked at 0−49 GPa (an average of the 0−49 GPa spectra). Vertical broken lines indicate olivine features.

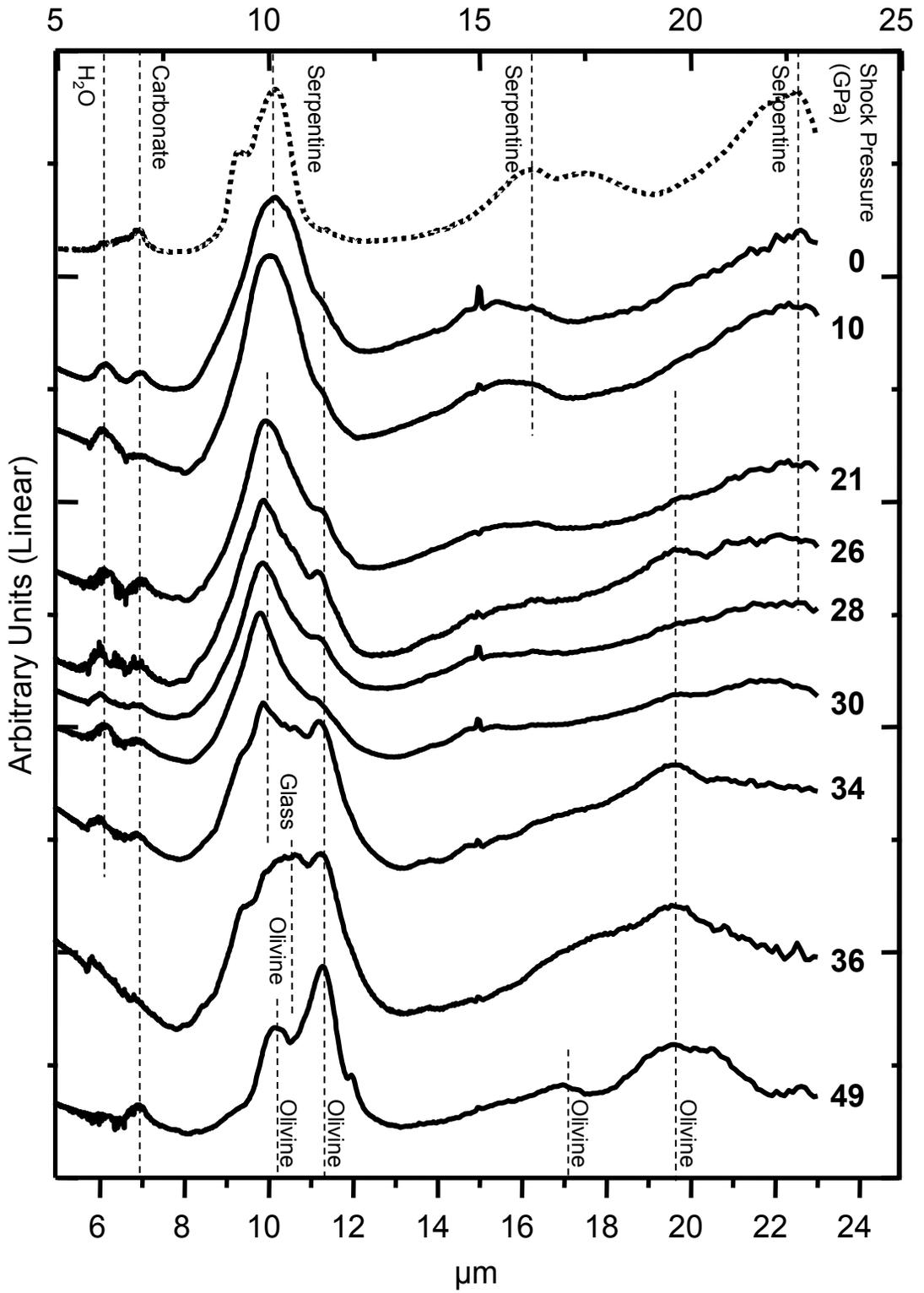

MORLOK Fig.1

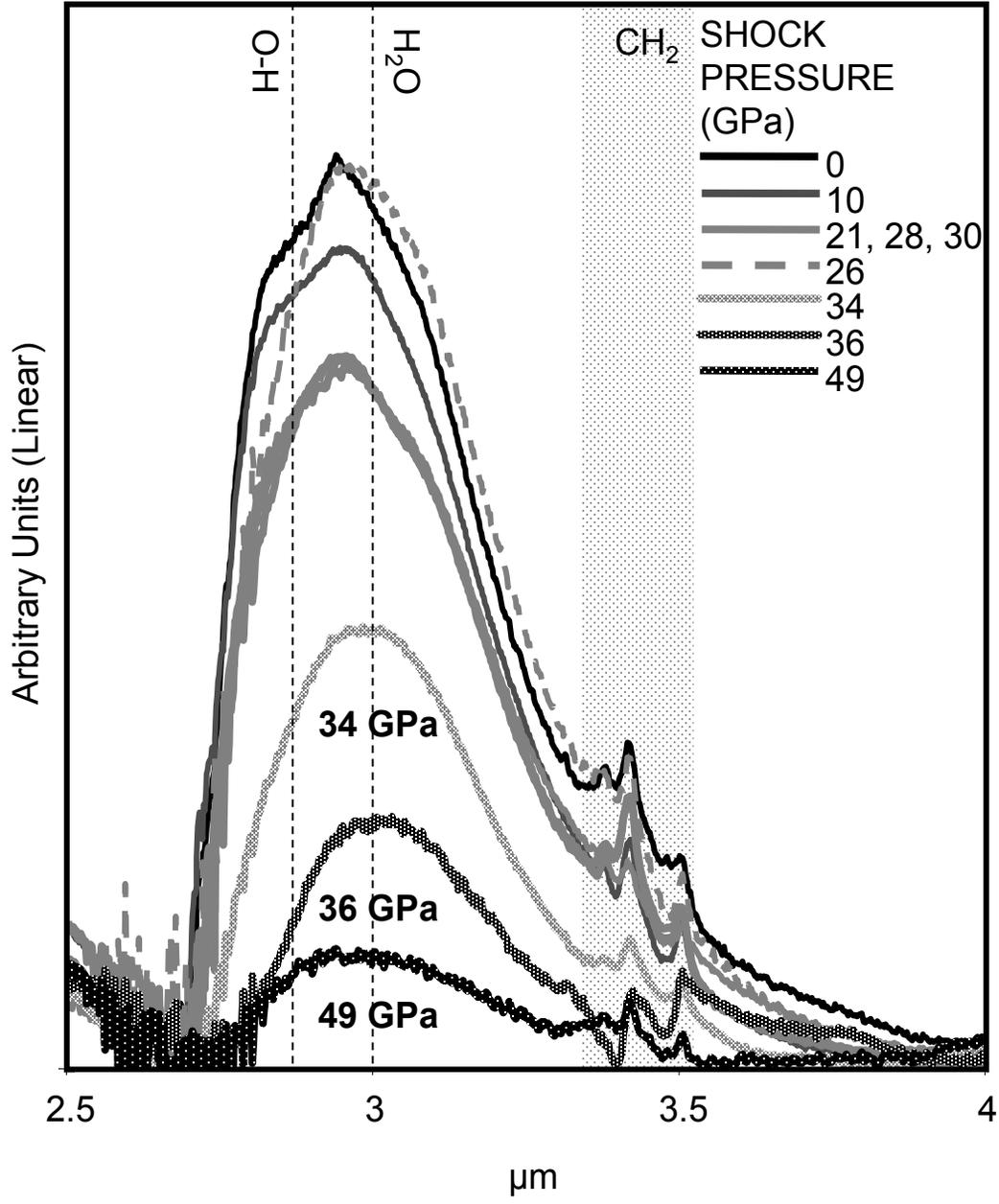

MORLOK Fig.2a

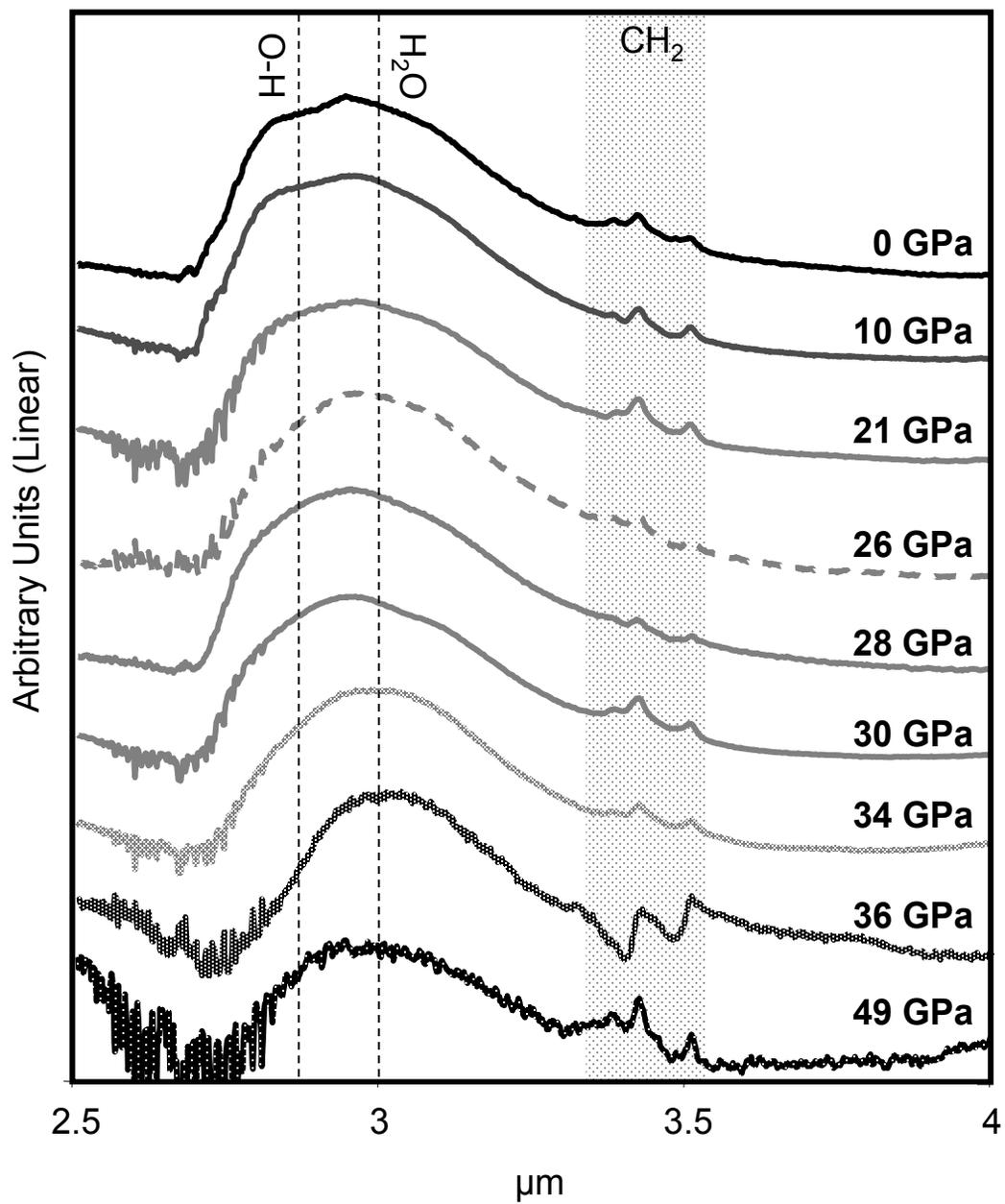

MORLOK Fig.2b

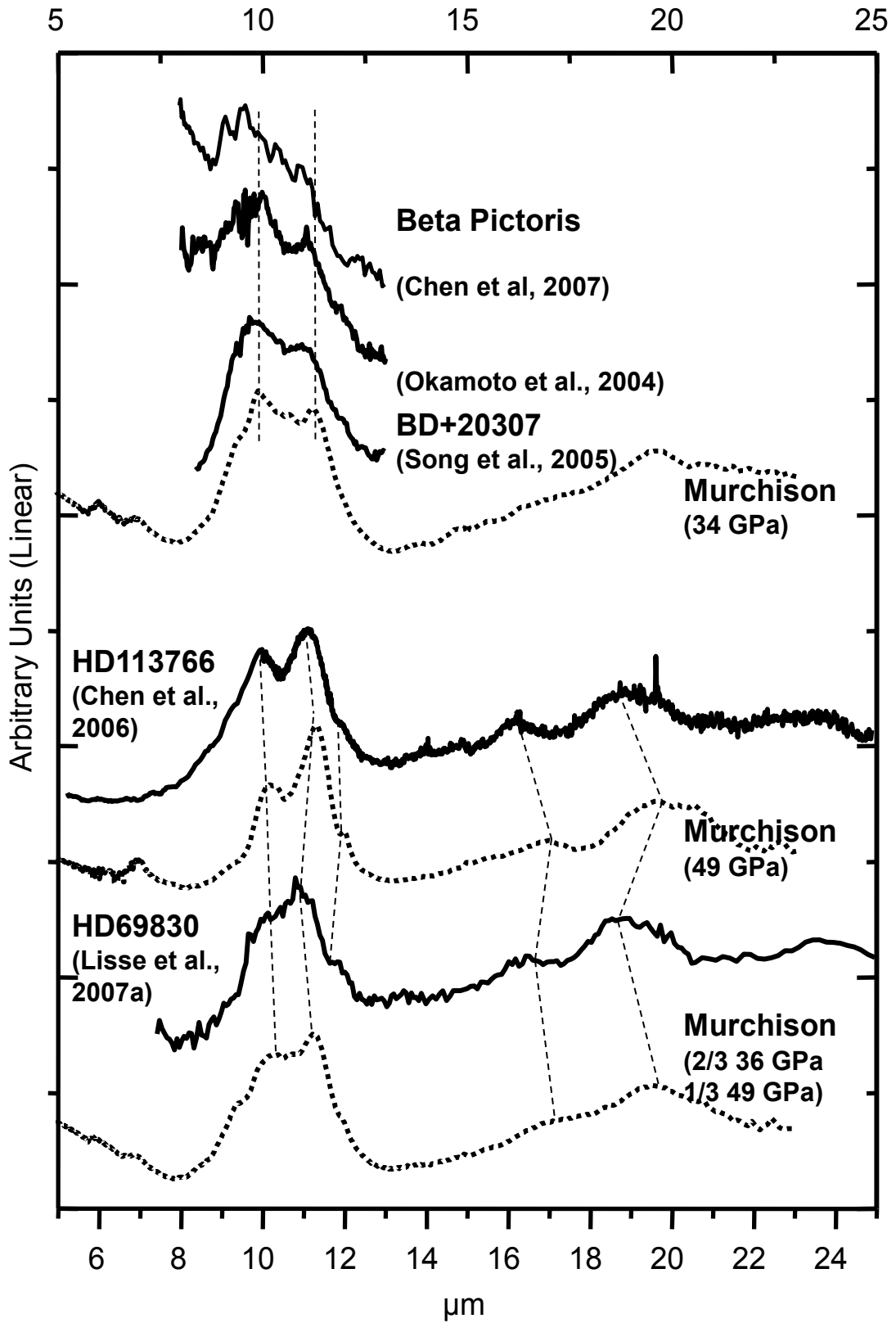

Beta Pictoris
(Chen et al, 2007)

(Okamoto et al., 2004)

BD+20307
(Song et al., 2005)

Murchison
(34 GPa)

HD113766
(Chen et al.,
2006)

Murchison
(49 GPa)

HD69830
(Lisse et al.,
2007a)

Murchison
(2/3 36 GPa
1/3 49 GPa)

MORLOK Fig.3

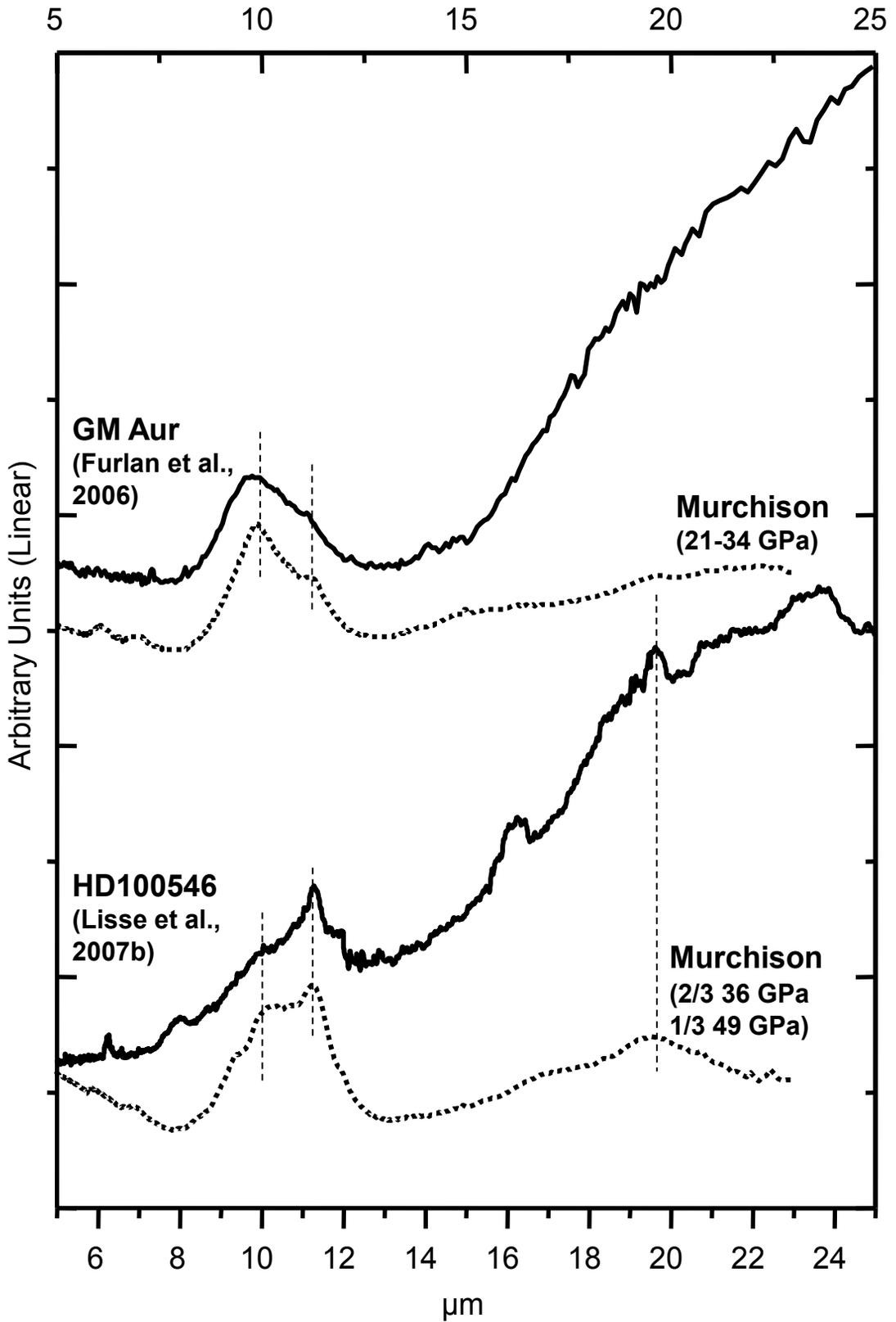

GM Aur
(Furlan et al.,
2006)

Murchison
(21-34 GPa)

HD100546
(Lisse et al.,
2007b)

Murchison
(2/3 36 GPa
1/3 49 GPa)

Arbitrary Units (Linear)

µm

MORLOK Fig.4

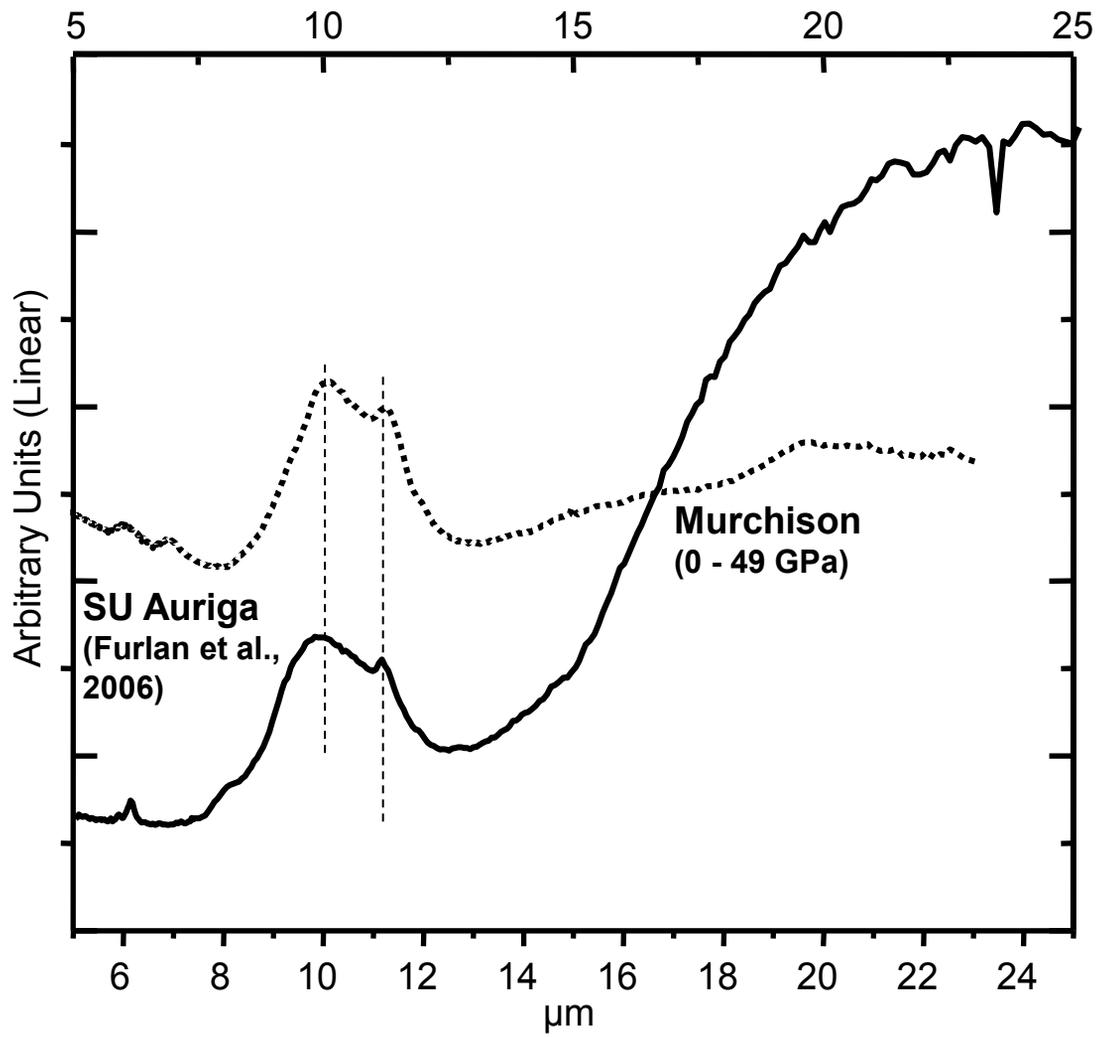

SU Auriga
(Furlan et al., 2006)

Murchison
(0 - 49 GPa)

MORLOK Fig.5

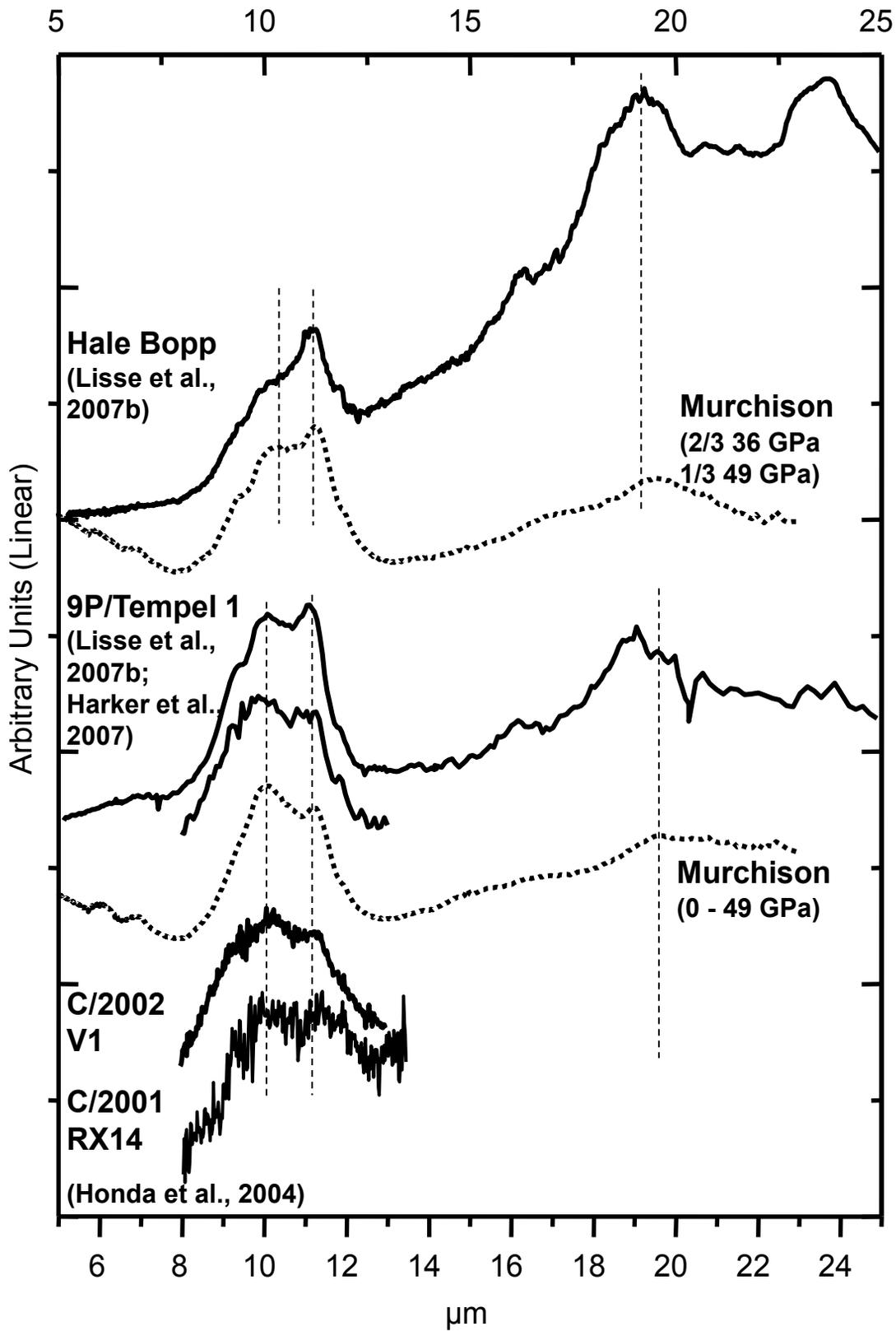

MORLOK Fig.6